\documentclass[journal]{IEEEtran}

\usepackage{cite}

\usepackage{color}

\usepackage{amsmath}

\usepackage{algorithm}
\usepackage{algorithmic}
\usepackage{bm}
\usepackage{latexsym}
\usepackage{amsthm}
\usepackage{url}
\usepackage{amsfonts}
\usepackage{amssymb}
\usepackage{indentfirst}
\usepackage{float}

\ifCLASSINFOpdf
   \usepackage[pdftex]{graphicx}
   \graphicspath{{../pdf/}{../jpeg/}}
   \DeclareGraphicsExtensions{.pdf,.jpeg,.png}
\else
   \usepackage[dvips]{graphicx}
   \graphicspath{{../eps/}}
   \DeclareGraphicsExtensions{.eps}
\fi
\usepackage{array}
\usepackage{multirow}

\ifCLASSOPTIONcompsoc
  \usepackage[caption=false,font=normalsize,labelfont=sf,textfont=sf]{subfig}
\else
  \usepackage[caption=false,font=footnotesize]{subfig}
\fi

\begin{document}
%
\title{A heterogeneous group CNN for image super-resolution}
%
%
%

\author{Chunwei Tian, \emph{Member}, \emph{IEEE},
        Yanning Zhang, \emph{Senior Member}, \emph{IEEE},
        Wangmeng Zuo, \emph{Senior Member}, \emph{IEEE},
        Chia-Wen Lin, \emph{Fellow}, \emph{IEEE},
        David Zhang, \emph{Life Fellow}, \emph{IEEE},
        Yixuan Yuan, \emph{Member}, \emph{IEEE} 
\thanks{This work was supported  in part by  the National Science Foundation of China under Grant 62201468, in part by the China Postdoctoral Science Foundation Grant 2022TQ0259, in part by the
Jiangsu Provincial Double–Innovation Doctor Program Under Grant JSSCBC20220942 and in part 
by the Shenzhen-Hong Kong Innovation Circle Category D Project
SGDX2019081623300177
. (Corresponding author: Yixuan Yuan (Email:  yxyuan@ee.cuhk.edu.hk)) }
\thanks{Chunwei Tian is with the School of Software, Northwestern Polytechnical University, Xi’an, Shaanxi, 710129, China. Also, he is with the National Engineering Laboratory for Integrated Aero-Space-Ground-Ocean Big Data Application Technology, Xi’an, Shaanxi, 710129, China.  (Email: chunweitian@nwpu.edu.cn)}
\thanks{Yanning Zhang is with the School of Computer Science, Northwestern Polytechnical University, the National Engineering Laboratory for Integrated Aero-Space-Ground-Ocean Big Data Application Technology, Xi’an, Shaanxi, 710129, China. (Email:ynzhang@nwpu.edu.cn)}
\thanks{Wangmeng Zuo is with the School of Computer Science and Technology, Harbin Institute of Technology, Harbin, Heilongjiang, 150001, China. (Email: wmzuo@hit.edu.cn)}
\thanks{Chia-Wen Lin is with the Department of Electrical Engineering and the Institute of Communications Engineering, National Tsing Hua University, Hsinchu, Taiwan. (Email: cwlin@ee.nthu.edu.tw)}
\thanks{David Zhang is with the School of Data Science, The Chinese University of Hong Kong (Shenzhen), Shenzhen, Guangdong, 518172, China. Also, he is with the Shenzhen Institute of Artificial Intelligence and Robotics for Society, Shenzhen, China. (Email: davidzhang@cuhk.edu.cn)}
\thanks{Yixuan Yuan is with the department of electronic engineering, The Chinese University of Hong Kong. (Email:  yxyuan@ee.cuhk.edu.hk)}}


\maketitle

\begin{abstract}
Convolutional neural networks (CNNs) have obtained remarkable performance via deep architectures. However, these CNNs often achieve poor robustness for image super-resolution (SR) under complex scenes. In this paper, we present a heterogeneous group SR CNN (HGSRCNN) via leveraging structure information of different types to obtain a high-quality image. Specifically, each heterogeneous group block (HGB) of HGSRCNN uses a heterogeneous architecture containing a symmetric group convolutional block and a complementary convolutional block in a parallel way to enhance internal and external relations of different channels for facilitating richer low-frequency structure information of different types. To prevent appearance of obtained redundant features, a refinement block with signal enhancements in a serial way is designed to filter useless information. To prevent loss of original information, a multi-level enhancement mechanism guides a CNN to achieve a symmetric architecture for promoting expressive ability of HGSRCNN. Besides, a parallel up-sampling mechanism is developed to train a blind SR model. Extensive experiments illustrate that the proposed HGSRCNN has obtained excellent SR performance in terms of both quantitative and qualitative analysis. Codes can be accessed at https://github.com/hellloxiaotian/HGSRCNN.
\end{abstract}


\section{Introduction}
\label{sec-1}
Single image super-resolution (SISR) aims to obtain more natural and realistic textures from a given low-resolution (LR) image to its high-resolution (HR) image, which is very beneficial to high-level tasks, i.e., image classification \cite{chen2021super} and object detection \cite{shermeyer2019effects}. Due to ill-pose inverse characteristic, SISR techniques have obtained enormous success via a degradation model with a priori knowledge, i.e., $L = {H_ \downarrow }_s$, where $L$ and $s$ represent a LR image and a scale factor, respectively \cite{zhang2018learning}. Also, $H$ denotes a predicted high-definition image. According to that, SISR methods can be summarized into three paradigms in general, i.e., interpolation methods, optimization methods, and discriminative learning methods. Interpolation methods mainly relied on bi-linear \cite{chiang1996efficient} or bicubic interpolation operations \cite{ha2018deep} to obtain a mapping from a LR image to a HR image. Although these methods were simple and efficient, they have obtained poor performance in SISR. To address this issue, optimization methods can be used to guide a SR model via natural image characteristics in a priori knowledge manner \cite{dong2012nonlocally}. For instance, using a sparse priori knowledge to obtain a linear combination can effectively predict HR images \cite{yang2010image}. However, this optimization method may enjoy a flexible work mode at the cost of a time-consuming process. Also, these methods may refer to manual setting parameters to achieve competitive SR performance. As an alternative, since discriminative learning methods have efficiency and flexibility, they are developed \cite{yang2019deep}. Notably, due to flexible end-to-end architectures, convolutional neural networks (CNNs) have dramatic demands in SISR \cite{wang2020deep}. The mentioned research can be generalized on two aspects in general, containing SR methods based high-frequency and low-frequency information. The SR methods based high-frequency information require size consistency of input and output in a CNN, which results in given LR images need be converted to high-frequency images through a bicubic operation as training images for constructing a SR model \cite{dong2015image}. Inspired by that, a very deep SR network architecture was implemented by using residual learning operations and stacking small filter sizes to obtain good visual effects \cite{kim2016accurate}. Due to deep architectures, CNNs are faced with training difficulty. To overcome the mentioned problem, recursive learning and residual learning techniques are presented to accelerate training speed \cite{kim2016deeply,tai2017image}. For instance, a deeply-recursive convolutional network (DRCN) integrated hierarchical information via residual learning techniques to facilitate accurate features for preventing exploding and vanishing gradients \cite{kim2016deeply}. Besides, fusing global and local information through skip connections to guide a new network architecture can enhance the learning ability for SISR \cite{mao2016image}. As an alternative, exploiting new components (i.e., recursive unit and gate unit) to obtain multi-level representation can improve the quality of a predicted image \cite{tai2017memnet}. Although these approaches can outperform traditional methods in SISR, they may refer to high complexity \cite{dong2016accelerating}. To overcome the challenge, SR methods based low-frequency information are developed. That is, directly inputting LR image into a CNN and using an up-sampling operation of deep layer to amplify obtained low-frequency features can train a SR model \cite{huang2021learning}. For example, designing a deformable and attentive mechanism to enhance a CNN extracted salient low-frequency texture information to enhance visual effects \cite{huang2021learning}. Although the methods above have achieved remarkable SR results, they only roughly fuse hierarchical features via residual learning or concatenation operations to affect different layers. That results in obtained features of simplification cannot represent well high-quality images, which achieves poor robustness in SISR under complex scenes.

In this paper, we propose a heterogeneous group SR CNN (HGSRCNN). It mainly uses heterogeneous group blocks (HGBs) to integrate structure information of different types for obtaining a HR image. Each HGB uses a heterogeneous architecture composing of a symmetric group convolutional block and a complementary convolutional block via enhancing internal and external relations of different channels in a parallel way to obtain more representative structure information of different types. Also, a refinement block with signal enhancement ideas in a serial way is developed to remove useless information for accelerating training efficiency. To alleviate loss of original information problem, a multi-level enhancement mechanism guides a CNN to construct a symmetric architecture for progressively facilitating information of HGSRCNN in SISR. Additionally, a parallel up-sampling mechanism is developed to train a blind SR model. 

 Main contributions of proposed HGSRCNN are conducted as follows.

(1)	The proposed 52-layer HGSRCNN uses heterogeneous architectures and refinement blocks to enhance internal and external interactions of different channels both in parallel and serial ways for obtaining richer low-frequency structure information of different types, which is very suitable to SISR in complex scenes. 

(2)	A multi-level enhancement mechanism guides a CNN to implement a symmetric architecture for progressively facilitating structural information in SISR.

(3) The designed HGSRCNN obtains competitive execution speed for SISR. That is, it only takes the run-time to $4.58\%$ of RDN and $4.27\%$ of SRFBN in restoring a high-quality image with $1024 \times 1024$. 

The remainder of this paper is conducted as follows. Section 2 reveals related work of the proposed method. Section 3 illustrates our proposed method. Section 4 gives experimental analysis and results. Section 5 concludes the proposed method. 
\section{Related work}
\subsection{Enhancement of different structure features for SISR}
Due to strong expressive ability, CNNs become popular in SISR. Notably, remarkable performance of CNNs is affected by deeper network architectures. To address this issue, enhancing structure features of deep networks can improve the interaction of both shallow and deep layers.  Mentioned techniques are usually classified into two categories: enhancements of high-frequency and low-frequency structure features. 

Enhancements of high-frequency structure features are composed of two stages \cite{tian2020lightweight}. The first stage utilizes bicubic-interpolation or bilinear-interpolation operations to zoom corrupted low-resolution images as high-frequency images. Then, using residual learning or concatenation operations integrates these high-frequency features via a designed CNN to facilitate richer structure features. Inspired by that, Kim et al. \cite{kim2016accurate} proposed a deeper network based on VGG via some small filters to obtain high-frequency structure information and a residual learning operation is used to enhance obtained structure information in SISR. Subsequently, a recursive CNN transferred structure features of shallow layers to the final layer through skip connections in a shared-parameter manner for enhancing clarity of predicted images \cite{kim2016deeply}. Alternatively, a multi-path residual CNN used global and local residual learning operations to fuse hierarchical structure features for improving a learning ability of a deep network in SISR \cite{tai2017image}. Besides, using skip-layer connections to connect multiple convolutional and deconvolutional layers for implementing a symmetrical network can also obtain more detailed structure features in SISR \cite{mao2016image}. Although these mentioned techniques have achieved excellent SR performance, they are faced with huge computational costs caused by training images of large sizes. To resolve this problem, enhancement methods of low-frequency structure features are presented. 

Enhancements of low-frequency structure features directly input corrupted low-resolution images into a CNN via a residual learning operation to extra robust low-frequency structure features, then they can use an up-sampling technique to deal with obtained low-frequency structure features for predicting high-quality images \cite{li2019feedback}. For instance, a residual dense network repeatedly used residual learning techniques to enhance effects of each layer for extracting more accurate low-frequency structure features in SISR \cite{zhang2018residual}. Alternatively, a multi-path residual network can aggregate different hierarchical features via different paths to enhance the robustness of obtained low-frequency structure information in SISR \cite{wang2021adversarial}. Besides, to accelerate the training efficiency, more refinement networks are presented \cite{ahn2018fast}. A cascade network with many smaller filters (i.e., convolutions of $1 \times 1$) utilized multiple shortcut connections to efficiently mine structure information of different types for obtaining a strong expressive ability of a SR model \cite{ahn2018fast}. Along this line, a coarse-to-fine SR CNN (CFSRCNN) applied residual learning and concatenation techniques in a heterogeneous architecture to respectively enhance low- and high-frequency structure information to improve the training stability and pursue excellent SR performance \cite{tian2020coarse}. The mentioned research shows that integrating different structure features is beneficial to SISR. Motivated by that, we design a diversified network architecture to enhance the effect of both internal connections from structure information of the same level and contextual structure information for SISR in this paper, according to training strategies of deep networks and signal processing knowledge.
\subsection{Deep CNNs based different channels for SISR}
Due to training difficultly of deeper network architectures, applying residual operation and skip connection techniques to transmit memory of shallow layers for obtaining more details of high-quality images are proposed \cite{yang2018drfn}. It is noted that these methods roughly fused hierarchical features to promote generalization abilities of SR models. They may cause a large computational burden. To address the phenomenon, deep CNNs with different channels are explored for SISR \cite{prajapati2021channel}. Referred methods can be roughly summarized into two categories: local channels and global channels.

The first strategy above splits all the channels via attention techniques to extract salient information for highlighting key channels in SISR, which can improve the training efficiency of SR models \cite{zhang2018image}. For instance, Zhang et al. \cite{zhang2018image} employed a residual channel attention mechanism to reinforce interdependencies of different channels, and adaptively filtered abundant low-frequency features for improving the performance of SISR. Besides, Niu et al. \cite{niu2020single} applied a holistic attention mechanism consisting of a layer attention module and a channel-spatial attention module to strengthen the correlations among different layers, channels and positions for selecting more expressive information in SISR.

The second mentioned strategy directly merged hierarchical channel information via residual learning and skip connection techniques to mine rich low-frequency features in SISR \cite{yang2020lightweight}. The simultaneous use of a dense network structure with group convolutions and small filters of $1 \times 1$ can enhance relationships of different channels via removing redundant parameters to progressively obtain useful information in SISR \cite{yang2020lightweight}. Along this line, Jain et al. \cite{jain2018efficient} unified group convolutional techniques and pruning ideas into a frame via throwing away useless information to reduce the test time of predicted HR images. Alternatively, aggregating obtained features from splitting convolutions by all the steps efficiently extracted discriminative information, i.e., edges, corners and textures to achieve clearer visual effects \cite{hui2019lightweight}. Besides, dividing a CNN into two sub-networks via a splitting operation to respectively learn robust hierarchical channel features can facilitate complementarity of different channels in MR image super-resolution \cite{zhao2019channel}. 

The research above illustrated that aggregating different information to enhance the interaction among different channels can achieve excellent SISR performance. Motived by that, we design a symmetric architecture via two twin branches to strengthen inner connections of different channels. Besides, we offer a supplementary block to learn features of all the channels, which can implement a complementary of both internal and external channels to obtain richer structure information in SISR. More information is given in Section 3. 
\section{Proposed Method}
The proposed HGSRCNN includes four components: two convolutional layers with rectified linear unit (ReLU) \cite{krizhevsky2012imagenet}, six symmetrical heterogeneous group blocks as well as HGBs, a parallel up-sampling mechanism and a single convolutional layer as illustrated in Fig.1. Specifically, each symmetrical heterogeneous group block uses a symmetric group convolutional and a complementary convolutional block in a parallel way to enhance internal and external relations of different channels for facilitating richer low-frequency structure information. Taking redundant features of the mentioned enhancement operation and training of deep CNNs into account, a refinement block with signal enhancement ideas is used into a HGB to remove useless information for accelerating the training. To prevent loss of original information, two enhancement branches are embedded into these HGBs to implement a local symmetrical architecture for progressively gathering low-frequency features in SISR. Besides, a parallel up-sampling with multiple scales is used to train a blind super-resolution model. Finally, a signal convolutional layer is employed to construct a HR image. More contents of the proposed HGSRCNN are given in latter parts. 
The proposed 52-layer incorporates 2-layer convolutions with ReLUs, 48-layer HGB, 1-layer parallel up-sampling mechanism and a single convolutional layer. The mentioned 2-layer convolutions with ReLUs are set as the 1st and 50th layers. Each convolution with a ReLU is equal to a convolutional layer acts a ReLU, which can be regarded as Conv+ReLU in Fig. 1. It can be observed that the first Conv+ReLU can obtain low-frequency features from an observation LR image through a convolutional operation, then obtained linear features can be mapped into a non-linearity via an activation function of a ReLU. Also, its parameters are set to input channels of 3, filter size of $3 \times 3$ and output channels of 64. Subsequently, six HGBs can extract richer low-frequency context structure information via enhancing internal and external relations of different channels in a parallel and serial way to obtain excellent SR performance. Parameters of each HGB are fixed as input channels of 64, filter size of $3 \times 3$ and output channels of 64, respectively. To prevent loss of original information, two enhancement branches are embedded into these HGBs to implement a local symmetrical architecture for progressively gathering low-frequency features in SISR, as described in Fig. 1. To avoid over-enhanced phenomenon from two enhancement branches above, the second Conv+ReLU is used to remove redundant low-frequency features, where its parameters are the same as HGB. That is, two enhancement branches (multi-level enhancement mechanism) act ends of both the first layer and the sixth HGB, ends of the second and fifth HGBs through residual learning operations, respectively. Besides, a parallel up-sampling mechanism can be exploited to map obtained low-frequency features into high-frequency features. It is noted that the referred to technique can simultaneously execute three different scales (i.e., $ \times 2$, $\times 3$ and $\times 4$) via a switch to train a blind model. Also, they enjoy the same setting as each HGB. Finally, a signal convolutional layer is conducted to obtain a predicted high-quality image through obtained high-frequency features. Its input and output channel number of 3 and filter size of $3 \times 3$ are given as parameters of the final layer. To conveniently understand work procedure of HGSRCNN, some characters are given. Let $P_{LR}$ and $P_{SR}$ express a given LR picture and a predicted HR picture of HGSRCNN, respectively. We assume that $C$ and $R$ be a convolutional operation and a function of ReLU, respectively. Also, $HGB$ is defined as function of a heterogeneous group block. $RL$ is regarded as a residual learning operation. Besides, $PUM$ denotes a parallel up-sampling mechanism. According to the motioned explanations, HGSRCNN can be expressed as

\begin{footnotesize}
\begin{equation}
\begin{array}{ll}
{P_{SR}} & = HGSRCNN({P_{LR}})\\
    &  = {C(PUM(R(C(G{E_2}{\rm{(}}HGB{\rm{(}}G{E_1}{\rm{(}}HG{B_5}{\rm{(}}{{\rm{O}}_1}{\rm{)))))))) }}},
\end{array}
\end{equation}
\end{footnotesize}

where $G{E_1}$ and $G{E_2}$ represent the first and second enhancement operations of HGB, respectively. Also, ${O_1} = R(C({P_{LR}}))$ . Besides, $HGSRCNN$ is function of HGSRCNN, which can be optimized via the following objective function. 
\subsection{Loss function}
To fairly optimize parameters of HGSRCNN, mean squared error (MSE) \cite{tian2020lightweight,douillard1995iterative} 
is selected as loss function to train a HGSRCNN model in SISR. HGSRCNN firstly uses a given LR image $P_{LR}$ as input of HGSRCNN to obtain predicted a HR image $P_{SR}$. Then, using MSE to compute the difference between obtained a HR image $P_{SR}$ and a given HR image $P_{HR}$ can optimize parameters. This process can be formulated as Eq. (2). 
\begin{equation}
\begin{array}{l}
LO(p) = \frac{1}{{2N}}\sum\limits_{j = 1}^N {{{\left\| {HGSRCNN(P_{LR}^j) - P_{HR}^j} \right\|}^2}},
\end{array}
\end{equation}
where $LO$ is loss function of MSE, $P_{LR}^j$ and $P_{HR}^j$ are the $jth$ LR and HR training images, respectively. Besides, $N$ denotes the number of training images. $p$ is treated as parameter set of training a HGSRCNN model. 

\subsection{Heterogeneous group block}
An 8-layer heterogeneous group block is used to facilitate more representative structure information of different types via a novel heterogeneous architecture to enhance internal and external relations of different channels
for improving SR performance and efficiency. Besides, to prevent redundancy of obtained features, a refinement block can further learn more accurate features. Designing signal enhancements fused into the refinement block can provide supplementary information of shallow layers for deep layers via integrating global and local low-frequency structure information in SISR. Detailed information of a heterogeneous group block is shown as follows. 

\begin{figure*}[!htbp]
\centering
\subfloat{\includegraphics[height=4.5in,width=5.5in]{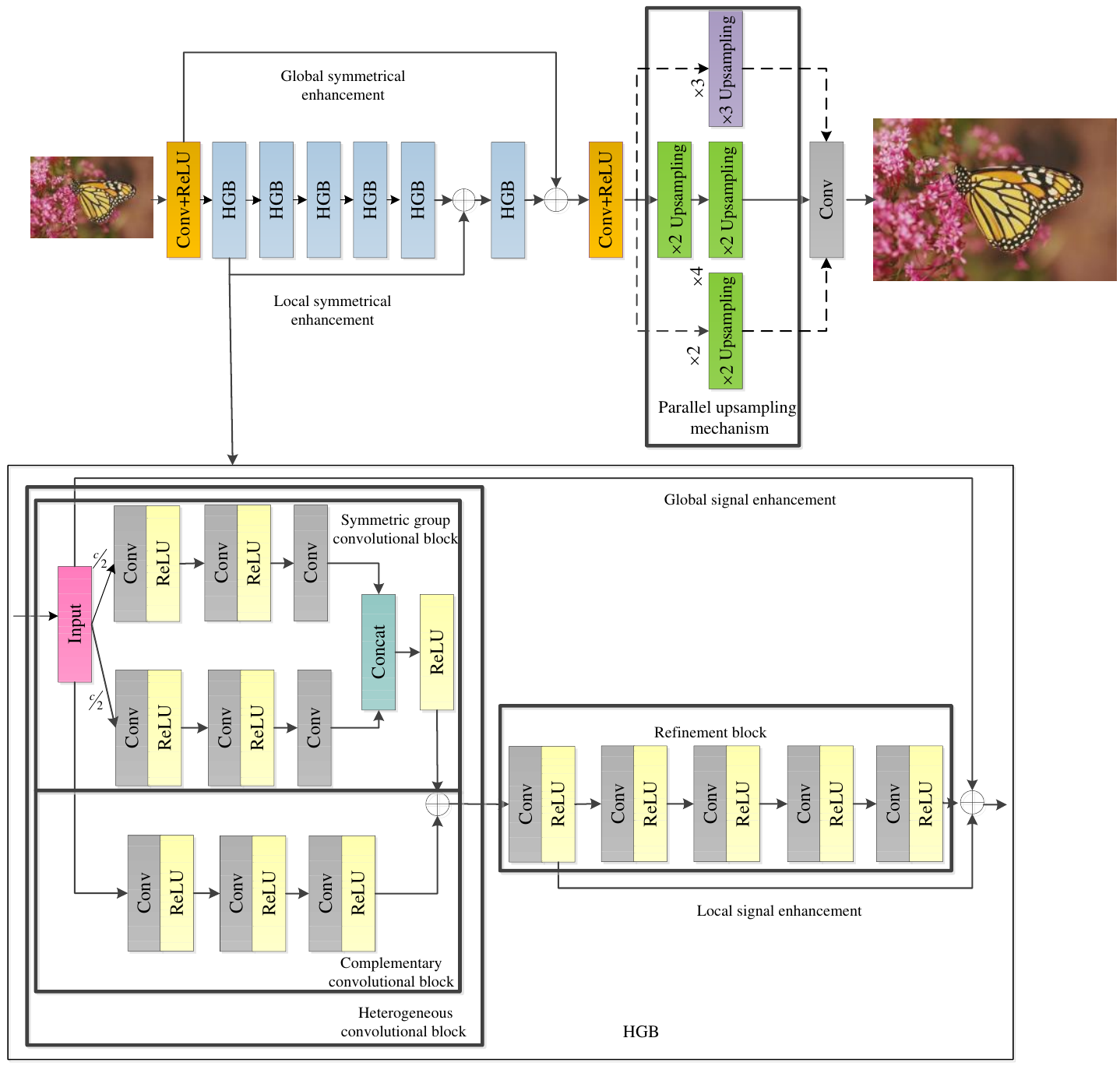}
\label{Fig:a}}
\hfil
\caption{Network architecture of HGSRCNN.}
\end{figure*}
\begin{small}
\end{small}
\begin{figure}[!htbp]
\centering
\subfloat{\includegraphics[width=3.5in]{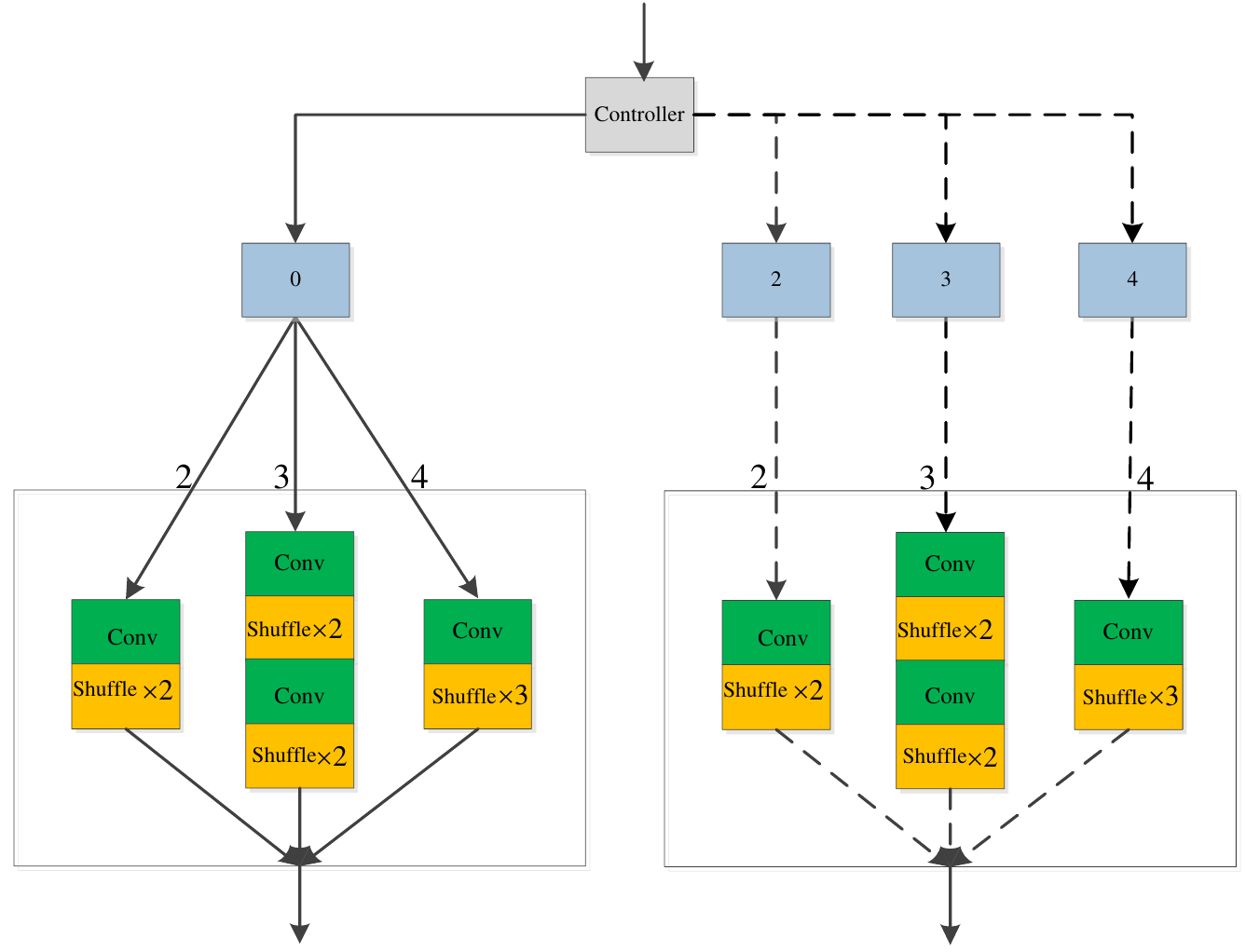}
\label{fig_second_case}}
\caption{Architecture of a parallel up-sampling mechanism.}
\label{fig:7}
\end{figure}

It is known that previous SR methods only directly fuse hierarchical features of all channels to enhance the SR performance, which may enhance the importance of redundant features to increase the convergence time of a SR model. To resolve this problem, we design heterogeneous group blocks via interacting different channels to extract wide and deep low-frequency structure information to enhance relation of different channels for improving the SR performance and efficiency. Specifically, each heterogeneous group block is composed of two parts: a heterogeneous convolutional block and a refinement block as shown in Fig. 1. 

heterogeneous convolutional block: The 3-layer heterogeneous convolutional block is composed of a symmetric group convolutional block and a complementary convolutional block is used to enhance internal and external relations of different channels for extracting robust low-frequency structure information. In terms of internal relation enhancement of different channels, two 3-layer sub-networks in the symmetric group convolutional block respectively learn representative information of split channels, integrate obtained features via a concatenation operation to enhance their internal correlations in SISR. Specifically, each layer of each sub-network is $Conv + ReLU$. Also, the input and output channels of each layer are 32, respectively. Their filter sizes are $3 \times 3$. Besides, output channel of symmetric group is 64, which is obtained by concatenating outputs of two sub-networks. To visually explain the mentioned process, the following formulas can be given.

Firstly, we use a splitting operation to divide input of current heterogeneous group 
block into two parts ($I_i^U$ and $I_i^P$) as inputs of two sub-networks in the symmetric group convolutional block as shown in Eq. (3) and Eq. (4), where $I_i^U$ and $I_i^L$ are the upper half and the lower half of all the channel features, respectively. 

\begin{small}
\begin{equation}
I_i^U = \left\{ {_{\frac{U}{2}{O_{i - 1}}}^{\frac{U}{2}{O_1}}{\rm{, }}} \right.
\end{equation}
\end{small}
\begin{small}
\begin{equation}
I_i^L = \left\{ {_{\frac{L}{2}{O_{i - 1}}}^{\frac{L}{2}{O_1}}{\rm{, }}} \right.
\end{equation}
\end{small}

where $O_{i - 1}$ denotes output of the $i-1th$ layer and $i >  = 2$. Specifically, $O_1$ expresses output of the first layer in the HGSRCNN. And, $O_{i - 1}$ denotes output of the $i - 2th$ heterogeneous group block. Also, $\frac{U}{2}$ and $\frac{P}{2}$ are defined as a splitting operation from channels of the upper half and the lower half, respectively. The obtained $I_i^U$ and $I_i^L$ act two sub-networks of a symmetric group convolutional block as illustrated shown Eq. (5).

\begin{small}
\begin{equation}
O_i^{SGCB} = R(Concat(C(R(C(R(C(I_i^U))))),C(R(C(R(C(I_i^L))))))),
\end{equation}
\end{small}

where $O_i^{SGCB}$ is output of symmetric group convolutional block in the $i - 1th$  HGB block ($2 =  < i <  = 7$) and $Concat$ denotes a concatenation operation as presented in Fig. 1. Also, output channel number of $O_i^{SGCB}$ is 64. 

Taking entirety of all channels into consideration, a 3-layer complementary convolutional block is designed to enhance their external correlations for enhancing the robustness of obtained features in SISR, which is complementary to symmetric group convolutional block. Each layer of complementary convolutional block is composed of $Conv + ReLU$. And parameters of each layer are input channel of 64, output channel of 64 and filter size of $3 \times 3$. Besides, output of a heterogeneous convolutional block can be obtained by a residual learning to fuse outputs of both symmetric group convolutional block and complementary convolutional block as an input of a refinement block. The procedure can be formulated as 

\begin{small}
\begin{equation}
O_i^{CCB} = R(C(R(C(R(C({I_i})))))),
\end{equation}
\end{small}
where $I_i$ is output of the upper layer. When the upper layer is the first layer, ${I_i} = {O_1}$. Otherwise, ${I_i}$ is $O_{i - 1}$ ($2 =  < i <  = 7$). Also, $O_i^{CCB}$ is output of a complementary convolutional block from the $i - 1th$  HGB. Subsequently, a residual learning operation is used to fuse outputs of a symmetric group convolutional block and a complementary convolutional block as an output of heterogeneous convolutional block as follows. 

\begin{small}
\begin{equation}
O_i^{HCB} = O_i^{SGCB} + O_i^{CCB},
\end{equation}
\end{small}
where $O_i^{HCB}$ expresses output of heterogeneous convolutional block in the $i - 1th$ HGB, which acts a refinement block. And $+$ denotes a residual learning operation, which is equal to $\oplus$ in Fig. 1.

A refinement block: To reduce importance of redundant information from heterogeneous convolutional block, a 5-layer refinement block is designed. Each layer of the refinement block is composed of $Conv + ReLU$ and their parameters are input channel of 64, output channel of 64 and filter size of $3 \times 3$. To strength the memory ability of shallow layers on deep layers in SISR, we use signal enhancement operations into the refinement block. That is, signal enhancement operations include a global signal enhancement and a local signal enhancement. The global signal enhancement utilizes a residual learning technique to fuse input of the heterogeneous convolutional block and output of the refinement block. The local signal enhancement utilizes a residual learning technique to integrate the output of the first layer in the refinement block and output of the refinement block. The implementation can be expressed as Eq. (8).

\begin{tiny}
\begin{equation}
\begin{array}{ll}
O_i^{HGB} & = R(C(R(C(R(C(R(C(R(C(O_i^{HCB})))))))))) + R(C(O_i^{HCB})) + {I_i}\\
       &  =  HGB{\rm{(}}O_i^{HCB}{\rm{)}},
\end{array}
\end{equation}
\end{tiny}
where $O_i^{HGB}$ is output of the $i - 1th$ HGB.  

\subsection{Multi-level enhancement mechanism}
To prevent loss of original information, a multi-level enhancement mechanism is embedded into these HGBs via two enhancement branches to implement a local symmetrical architecture for progressively gathering low-frequency features in SISR, as described in Fig. 1. The first enhancement branch (global symmetrical enhancement) is that fuses the outputs of the first HGB and fifth HGB via a residual learning operation as  input of the sixth HGB. The mentioned implementations can be described as follows. 

\begin{small}
\begin{equation}
\begin{array}{ll}
{I_6} & = G{E_1}(HG{B_5}({O_1}))\\
       &  =  O_5^{HGB} + O_2^{HGB},
\end{array}
\end{equation}
\end{small}
where $G{E_1}$ expresses a function of the first enhancement branch and $HG{B_5}$ is symbolled as functions of five HGBs. ${I_6}$ is input of the 6th HGB. Besides, $O_5^{HGB}$ and $O_2^{HGB}$ denote outputs of the 2rd and 5th HGBs, respectively. To further improve importance of hierarchical features, the second enhancement branch (local symmetrical enhancement) is designed by a residual learning operation. The second enhancement branch acts both the 1st layer of HGSRCNN and the 6th HGB as follows. 

\begin{small}
\begin{equation}
\begin{array}{ll}
{O_{HGBS}} & = G{E_2}(O_6^{HGB})\\
       &  =  {O_1} + O_6^{HGB},
\end{array}
\end{equation}
\end{small}
where $O_{HGBS}$ is output of all the HGBs as input of the second $C{\rm{onv + ReLU}}$. The $G{E_2}$ denotes function of the second enhancement branch. The second $Conv + ReLU$ is used to prevent the over-enhancement phenomenon of HGBs and it acts a parallel up-sampling mechanism. 

\subsection{Parallel up-sampling mechanism}
Due to ill-posed inverse characteristic of image super-resolution, scholars tend to 
establish a SR model via a certain scale. However, LR images have suffered from different corruption, which makes most of existing SR models cannot exert effects \cite{agustsson2017ntire}. To resolve this issue, a parallel up-sampling mechanism \cite{ahn2018fast} with a flexible controller is used in the HGSRCNN to achieve a blind super-resolution model. Its implementations and work mechanism can be illustrated as follows. 

The parallel up-sampling mechanism contains three components, i.e., $ \times 2$ $Upsampling$,  $ \times 3$ $Upsampling$ and $\times 4$ $Upsampling$. Specifically, $ \times 2$ $Upsampling$,  $ \times 3$ $Upsampling$ and $\times 4$ $Upsampling$ can be respectively equal to a Conv+Shuffle $ \times 2$, Conv+Shuffle $ \times 3$ and Conv+Shuffle $ \times 4$  (also regarded as two Conv+Shuffle $ \times 2$ ) ,  where $ \times 2$  $Upsampling$ and $ \times 3$ $Upsampling$ denote a convolution with size of $3 \times 3$ and Shuffle $\times 2$ and Shuffle $\times 3$, respectively. Also, input and output channels of each component are 64. Besides, a flexible controller can control different components to obtain a blind super-resolution model. That is, if the controller value is 0, three components will parallel work to train a SR model for different scales (i.e., $\times 2$, $\times 3$ and $\times 4$) as presented in Fig. 2, which is expressed by mentioned solid line part. Otherwise, the controller value can be extended to be a scale from 2, 3 and 4, a super-resolution model with a certain scale is obtained, which is represented by mentioned dotted line part in Fig. 2. To intuitively show the execution process, the following equation is conducted.

\begin{tiny}
\begin{equation}
\begin{array}{ll}
{O_{PUM}} & = PUM({O_{HGBS}})\\
         & = \{ _{P{S_i}{(}C{(O}_{HGBS}^i{))                                                                  , }i = 2,3,4}^{P{S_2}{(}C{(O}_{HGBS}^2{))} \circ P{S_3}{(}C{(O}_{HGBS}^3{))} \circ P{S_2}(C(P{S_2}{(}C{(O}_{HGBS}^4{)))),} i = 0 }
\end{array}
\end{equation}
\end{tiny}

where $O_{PUM}$ denotes output of the parallel up-sampling mechanism. $O_{HGBS}^2$, $O_{HGBS}^3$ and $O_{HGBS}^4$ are used to stand for outputs of obtained low-frequency structure information for $ \times 2$ ,  $ \times 3$ and $\times 4$, respectively. Also, $P{S_2}$, $P{S_3}$ and $P{S_4}$ are symbolized as the functions of $\times 2$ $Upsampling$,  $\times 3$ $Upsampling$ and $\times 4$ $Upsampling$, respectively.  Let $\circ$  express a parallel operation. $O_{HGBS}^i$  and $P{S_i}$ are used to represent low-frequency output and function of an up-sampling operation for a scale factor with $i$, respectively. Besides, $O_{PUM}$ acts a single convolutional layer as the last layer in the HGSRCNN as given in Eq. (12), which can be utilized to construct predicted high-quality images. Its parameters are input channel number of 64, output channel number of 3 and filter size of $3 \times 3$. 
\begin{equation}
\begin{array}{ll}
   {P_{SR}} = C({O_{PUM}})
\end{array}
\end{equation}
\section{Experiments}
\subsection{Training datasets}
To guarantee experimental fairness, a popular color image dataset of DIV2K \cite{agustsson2017ntire} is used to train a HGSRCNN model. The DIV2K contains training samples of 800 natural images, validation samples of 100 natural images and test samples of 100 natural images for different scales in $ \times 2$, $ \times 3$  and $ \times 4$. Besides, to make obtained SR model more robust, the following data augment way is exploited to enlarge the training dataset \cite{tian2020coarse}. Firstly, the training dataset and validation dataset from the same scale are merged into a new training dataset for training a HGSRCNN model. Secondly, to improve the training efficiency of a HGSRCNN model, each LR image is cropped as patches of size $81 \times 81$. Finally, random horizontal flips and rotation operation of $90^\circ$ are used to deal with these patches for extending categories of training samples. 

\subsection{Testing datasets}
Inspired by popular SR methods (i.e., LESRCNN \cite{tian2020lightweight}, CARN \cite{ahn2018fast} and CFSRCNN  \cite{tian2020coarse}), four public datasets containing Set5 \cite{bevilacqua2012low}, Set14 \cite{bevilacqua2012low}, BSD100 (B100) \cite{martin2001database} and Urban100 (U100) \cite{huang2015single} of $\times 2$, $\times 3$ and $\times 4$  are used as test datasets. The Set5 and Set14 have respectively captured five and fourteen color images via same digital devices for three scales ($\times 2$,  $\times 3$ and $\times 4$). B100 and U100 include a hundred color images for $\times 2$, $\times 3$ and $\times 4$, respectively.

Motivated by state-of-the art SR methods \cite{tian2020lightweight,ahn2018fast}, Y channel in YCbCr space is chosen to conduct experiments in this paper. That is, predicted RGB images of a HGSRCNN model need be converted as images of Y channel to test the performance of a designed HGSRCNN for image super-resolution. 

\subsection{Experimental settings}
To better train a blind model, initial parameters are given as follows. Initial learning rate is set to le-4, which may be halved for every 4e+5 steps from 553,000 steps. Also, batch size is treated as 32, epsilon of 1e-8, $\beta _1$ of 0.9, $\beta _2$ of 0.999 and more initial parameters are referred to Refs.\cite{tian2020lightweight,ahn2018fast}. Also, the controller value of 0 is conducted experiments in this paper.  Besides, training parameters are updated via an optimizer of Adam \cite{kingma2014adam}. 

The HGSRCNN network is implemented by Pytorch of 1.2.0, Python of 3.6.6 on a Ubuntu system of 16.04. Besides, a PC containing RAM of 16G, one graphic processing unit (GPU) \cite{bergstra2010theano} with Inter Core i7-7800 and two GPUs with Nvidia GeForce GTX 1080Ti is used to provide computational ability. Specifically, mentioned GPUs reply on Nvidia CUDA of 11.3 and cuDNN of 8.0 to improve the execution speed.
\subsection{Network analysis}
As is known to all, merging hierarchical features can enhance the importance of shallow layers on deep layers to promote the SR performance \cite{tai2017memnet}. However, most of these SR techniques roughly merge obtained features of all the channels rather than strengthening the effects of local salient channels, which may result in obtained information of simplification cannot completely express high-quality images and achieve poor robustness for SISR of complex scenes. To address this issue, we present a heterogeneous group SR CNN (HGSRCNN) via integrating structure information of different types to enhance relations of different channels. Specifically, a heterogeneous group block (HGB) uses a symmetric group convolutional and a complementary convolutional block to enhance internal and external relations of different channels to obtain more expressive structure information of different types. Also, a refinement block with signal enhancement ideas is fused into a heterogeneous group block to filter useless information for accelerating training efficiency. To alleviate original information loss problem, a multi-level enhancement mechanism guides a CNN to construct a symmetric architecture via different HGBs and residual learning operations for progressively facilitating information of HGSRCNN in SISR. Additionally, a parallel up-sampling mechanism is developed to train a blind SR model. More detailed information of HGSRCNN in design principle of a network architecture and effectiveness of important components are descripted as follows. 

HGSRCNN contains two Conv+ReLU, several heterogeneous group blocks (HGBs), a parallel up-sampling mechanism and a single Conv. The first Conv+ReLU can be employed to convert a given low-resolution image into non-linear low-frequency features. According to VGG architecture \cite{simonyan2014very}, increasing the depth of network can mine more useful features. Inspired by that, six HGBs are stacked behind the first Conv+ReLU to extract richer low-frequency structure information. The design rules of each HGB are conducted via enhancing relations of different channels and training strategies of deep networks. 

Relations of enhancing different channels: To enhance the expressive ability, some methods only roughly fuse hierarchical features through a residual learning or concatenation operations to strengthen effects of different layers. However, due to simplification of obtained features, they cannot represent well high-quality images, which obtained poor robustness in SISR under complex scenes. To address this problem, we design a heterogeneous architecture via enhancing  internal and external relations of different channels to extract more accurate low-frequency features. 

In terms of enhancing internal relations of different channels, we propose a 3-layer symmetric group convolutional block. The mentioned symmetric group convolutional block firstly halves output channels of the last HGB into two parts as inputs of two sub-networks. Next, each sub-network is used to learn more accurate low-frequency channel structure information, respectively. Finally, a concatenation operation is used to merge obtained features from two sub-networks. Although the mentioned mechanism can enhance the internal relations of different channels, they ignore overall of obtained features from all the channels. Increasing the width of a deep network can capture more complementary information, according to the GoogLeNet \cite{szegedy2015going}.  Inspired by that, we design a 3-layer complementary convolutional block to strengthen external relations of different channels for mining complementary low-frequency struct features, which makes an obtained super-resolution model robust for complex scenes. Because the complementary convolutional block and symmetric group convolutional block are parallel executed, the process is treated as a parallel procedure. Besides, we conduct a TABLE I to verify effectiveness of the mentioned blocks. That is, a 7-layer normal convolutional network (NCN) can be better than a symmetric group convolutional network (SGCN) in Peak signal-to-noise ratio (PSNR) \cite{hore2010image} and structural similarity index (SSIM) \cite{hore2010image} on U100 for
$\times 2$, where NCN denotes a combination of 5-layer Conv+ReLU, 1-layer parallel upsampling mechanism, 1-layer Conv and SGCN denotes a combination of 3-layer symmetric group convolutional block, 1-layer Conv+ReLU, 1-layer parallel upsampling mechanism and 1-layer Conv. Also, the number of parameters from the SGCN is $71\%$ of the NCN as shown in TABLE II. In a summary, the proposed symmetric group convolutional block can make a tradeoff between performance and complexity. Besides, the HSRCNN without GSE, LSE, LOSE and refinement block (RB) has remarkable improvement than that of HGSRCNN without GSE, LSE, LOSE, RB and complementary convolutional block (CCB) in both PSNR and SSIM on U100 for 
$\times 2$ as shown in TABLE I, where GSE and LSE denote a global symmetrical enhancement and a local symmetrical enhancement, respectively. Also, the HGSRCNN without GSE, LSE, LOSE, RB is more superior than that of NCN on U100 for $\times 2$ as illustrated in TABLE I, where RB expresses a refinement block and LOSE denotes a local signal enhancement. That shows that the combination of the proposed symmetric group convolutional block and complementary convolutional block is more effective in SISR. Although the heterogeneous convolutional block can enhance the relations of different channels, they may include redundant information to affect the training speed. 
\begin{table*}[htbp!]
\caption{PSNR and SSIM of different SR methods on U100 for $\times 2$.}
\label{tab:1}
\centering
\scalebox{0.8}[0.75]{
\begin{tabular}{|c|c|c|}
\hline
Methods &PSNR (dB) &SSIM\\
\hline
Normal convolutional network (NCN)  &30.59	&0.9121\\
\hline
Symmetric group convolutional network (SGCN)	&30.42 &0.9088\\
\hline
HGSRCNN without GSE, LSE, LOSE, RB and complementary convolutional block (CCB) &31.23 &0.9186\\
\hline
HGSRCNN without GSE, LSE, LOSE and refinement block (RB) &31.74 &0.9239\\
\hline
HGSRCNN without GSE, LSE and local signal enhancement (LOSE) &32.15 &0.9285\\
\hline
HGSRCNN without LSE, and global symmetrical enhancement (GSE) &32.17 &0.9288\\
\hline
HGSRCNN without local symmetrical enhancement (LSE)  &32.20 &0.9286\\
\hline
HGSRCNN (Ours) &32.21 &0.9292\\
\hline
\end{tabular}}
\label{tab:booktabs}
\end{table*}

\begin{table}[htbp!]
\caption{Complexity of different SR networks.}
\label{tab:1}
\centering
\scalebox{0.80}[0.85]{
\begin{tabular}{|c|c|c|}
\hline
Methods &Parameters &Flops\\
\hline
SGCN  &132.48K	&1.63G\\
\hline
NCN	&187.78K	&1.99G\\
\hline
\end{tabular}}
\label{tab:booktabs}
\end{table}

Refinement block: To resolve this problem, we propose a refinement block to learn more accurate low-frequency structure information in a serial way. Also, increasing the depth of a deep network can enlarge receptive field to mine more useful information, according to VGG\cite{simonyan2014very}. Motived by that, a stacked 5-layer Conv+ReLU forms a refinement block. Its effectiveness is proved by HGSRCNN without GSE, LSE, LOSE and refinement block (RB) and HGSRCNN without GSE, LSE and local signal enhancement (LOSE) in TABLE I. Besides, it is known that the depth of a deep network is bigger, its performance may drop \cite{tai2017memnet}. To tackle this problem, signal enhancement operations are gathered in a HGB.

The mentioned signal enhancement operations depend on two signal enhancements to strengthen the memory abilities of shallow layers on deep layers in SISR, according to training strategies of deep networks. That is, signal enhancement operations include a global signal enhancement (GOSE) and a local signal enhancement as well as LOSE. The GOSE utilizes a residual learning technique to fuse input of the heterogeneous convolutional block and output of the refinement block. The LOSE utilizes a residual learning technique to integrate output of the first layer in the refinement block and output of the refinement block. Also, their effective results are verified as presented in TABLE I. That is, HGSRCNN without LSE, and global symmetrical enhancement outperforms HGSRCNN without GSE, LSE and local signal enhancement in both PSNR and SSIM on U100 for $\times 2$. Although mentioned HGBs can effectively mine low-frequency information, they ignore relations of different HGBs. To prevent the phenomenon, a multi-level enhancement mechanism is designed as follows.

Multi-level enhancement mechanism: This mechanism relies on two enhancement branches to make HGSRCNN implement a local symmetrical architecture for progressively gathering low-frequency features for SISR as described in Fig.1. The first enhancement branch (local symmetrical enhancement) is that fuses the outputs of the first HGB and fifth HGB via a residual learning operation as input of the sixth HGB. Its effect is tested via comparisons between HGSRCNN and HGSRCNN without local symmetrical enhancement in Table I. The second enhancement branch as well as global symmetrical enhancement is used to fuse outputs of the 1st layer of HGSRCNN and the 6th HGB. From TABLE I, we can see that HGSRCNN with local symmetrical enhancement can obtain better results than that of HGSRCNN without LSE and global symmetrical enhancement, which also implies importance of the GSE in SISR. Additionally, to avoid the over-enhanced phenomenon from two enhancement branches above, a Conv+ReLU is used to remove redundant low-frequency features, where its parameters are the same as input channel number, output channel number and filter size of each HGB. To handle blind super-resolution, a parallel upsampling mechanism is used to implement a super-resolution for multiple scales as shown in Eq. (12). Besides, a convolutional layer is used as the last layer to obtain predicted high-quality images. 

\begin{table}[htbp!]
\caption{Average PSNR/SSIM results of different SR techniques for different scales ($\times 2$, $\times 3$ and $\times 4$) on Set5.}
\label{tab:1}
\centering
\scalebox{0.75}[0.75]{
\begin{tabular}{|c|c|c|c|c|}
\hline
\multirow{2}{*}{Dataset} &
\multirow{2}{*}{Methods} &
$\times 2$ & $\times 3$ & $\times 4$\\
\cline{3-5} & &PSNR/SSIM &PSNR/SSIM &PSNR/SSIM\\
\hline
\multirow{38}{*}{Set5} &
Bicubic\cite{sun2008image}	&33.66/0.9299	&30.39/0.8682	&28.42/0.8104\\
\cline{2-5} &
A+\cite{timofte2014a+}	&36.54/0.9544	&32.58/0.9088	&30.28/0.8603\\
\cline{2-5} &
RFL \cite{schulter2015fast}	&36.54/0.9537	&32.43/0.9057	&30.14/0.8548\\
\cline{2-5} &
SelfEx\cite{huang2015single}	&36.49/0.9537 &32.58/0.9093 &30.31/0.8619\\
\cline{2-5} &
CSCN\cite{wang2015deep} &36.93/0.9552 &33.10/0.9144 &30.86/0.8732\\
\cline{2-5} &
RED30\cite{mao2016image} &37.66/0.9599 &33.82/0.9230	&31.51/0.8869\\
\cline{2-5} &
DnCNN\cite{zhang2017beyond}	&37.58/0.9590	&33.75/0.9222 &31.40/0.8845\\
\cline{2-5} &
TNRD\cite{chen2016trainable} &36.86/0.9556	&33.18/0.9152 &30.85/0.8732\\
\cline{2-5} &
FDSR\cite{lu2018fast} &37.40/0.9513	&33.68/0.9096 &31.28/0.8658\\
\cline{2-5} &
SRCNN\cite{dong2015image}	&36.66/0.9542	&32.75/0.9090 &30.48/0.8628\\
\cline{2-5} &
FSRCNN\cite{dong2016accelerating} &37.00/0.9558	&33.16/0.9140 &30.71/0.8657\\
\cline{2-5} &
RCN\cite{shi2017structure} &37.17/0.9583	&33.45/0.9175	&31.11/0.8736\\
\cline{2-5} &
VDSR\cite{kim2016accurate} &37.53/0.9587	&33.66/0.9213	&31.35/0.8838\\
\cline{2-5} &
DRCN\cite{kim2016deeply} &37.63/0.9588	&33.82/0.9226	&31.53/0.8854\\
\cline{2-5} &
CNF\cite{ren2017image}	&37.66/0.9590	&33.74/0.9226	&31.55/0.8856\\
\cline{2-5} &
LapSRN\cite{lai2017deep} &37.52/0.9590	&-	&31.54/0.8850\\
\cline{2-5} &
IDN\cite{hui2018fast}  &37.83/\textcolor{blue}{0.9600} &34.11/0.9253 &31.82/0.8903\\
\cline{2-5} &
DRRN\cite{tai2017image} &37.74/0.9591  &34.03/0.9244 &31.68/0.8888 \\
\cline{2-5} &
BTSRN\cite{fan2017balanced} &37.75/- &34.03/- &31.85/-\\
\cline{2-5} &
MemNet\cite{tai2017memnet}  &37.78/0.9597  &34.09/0.9248 &31.74/0.8893\\
\cline{2-5} &
CARN-M\cite{ahn2018fast}  &37.53/0.9583 &33.99/0.9236 &31.92/0.8903\\
\cline{2-5} &\
CARN\cite{ahn2018fast}  &37.76/0.9590	&\textcolor{blue}{34.29}/0.9255	&\textcolor{red}{32.13}/\textcolor{blue}{0.8937}\\
\cline{2-5} &\
EEDS+\cite{wang2019end} &37.78/\textcolor{red}{0.9609} &33.81/0.9252 &31.53/0.8869\\
\cline{2-5} &
DRFN\cite{yang2018drfn} &37.71/0.9595 &34.01/0.9234 &31.55/0.8861\\
\cline{2-5} &
MSDEPC\cite{liu2019single} &37.39/0.9576 &33.37/0.9184 &31.05/0.8797\\
\cline{2-5} &
CFSRCNN\cite{tian2020coarse} &\textcolor{blue}{37.79}/0.9591	&34.24/\textcolor{blue}{0.9256}	&\textcolor{blue}{32.06}/0.8920\\
\cline{2-5}&
LESRCNN\cite{tian2020lightweight} &37.65/0.9586	&33.93/0.9231	&31.88/0.8903\\
\cline{2-5}&
LESRCNN-S\cite{tian2020lightweight} 	&37.57/0.9582	&34.05/0.9238	&31.88/0.8907\\
\cline{2-5} &
ACNet\cite{tian2021asymmetric} 	&37.72/0.9588	&34.14/0.9247	&31.83/0.8903\\
\cline{2-5} &
ACNet-B\cite{tian2021asymmetric}  &37.60/0.9584	&34.07/0.9243	&31.82/0.8901\\
\cline{2-5} &
DIP-FKP\cite{liang2021flow} &30.16/0.8637 &28.82/0.8202 &27.77/0.7914\\
\cline{2-5} &
DIP-FKP + USRNet\cite{liang2021flow} &32.34/0.9308 &30.78/0.8840 &29.29/0.8508\\
\cline{2-5} &
KOALA\cite{kim2021koalanet} &33.08/0.9137 &- & 30.28/0.8658\\
\cline{2-5} &
FALSR-B\cite{chu2021fast} &37.61/0.9585 &- &-\\
\cline{2-5} &
FALSR-C\cite{chu2021fast} &37.66/0.9586 &- &-\\
\cline{2-5} &
ESCN \cite{wang2017ensemble} &37.14/0.9571 &33.28/0.9173 &31.02/0.8774\\
\cline{2-5} &
HDN \cite{jiang2020hierarchical} &37.75/0.9590 &34.24/0.9240 &\textcolor{red}{32.23/0.8960}\\
\cline{2-5} &
HGSRCNN (Ours)	&\textcolor{red}{37.80}/0.9591	&\textcolor{red}{34.35/0.9260}	&\textcolor{blue}{32.13/0.8940}\\
\hline
\end{tabular}}
\label{tab:booktabs}
\end{table}
\begin{table}[t!]
\caption{Average PSNR/SSIM results of different SR techniques for different scales ($\times 2$, $\times 3$ and $\times 4$) on Set14.}
\label{tab:1}
\centering
\scalebox{0.75}[0.75]{
\begin{tabular}{|c|c|c|c|c|}
\hline
\multirow{2}{*}{Dataset} &
\multirow{2}{*}{Methods} &
$\times 2$ & $\times 3$ & $\times 4$\\
\cline{3-5} & &PSNR/SSIM &PSNR/SSIM &PSNR/SSIM\\
\hline
\multirow{38}{*}{Set14} &
Bicubic\cite{sun2008image}	&30.24/0.8688	&27.55/0.7742	&26.00/0.7027\\
\cline{2-5} &
A+\cite{timofte2014a+}	&32.28/0.9056	&29.13/0.8188	&27.32/0.7491\\
\cline{2-5} &
RFL\cite{schulter2015fast}	&32.26/0.9040	&29.05/0.8164	&27.24/0.7451\\
\cline{2-5} &
SelfEx\cite{huang2015single}	&32.22/0.9034 &29.16/0.8196	&27.40/0.7518\\
\cline{2-5} &
CSCN\cite{wang2015deep} &32.56/0.9074	&29.41/0.8238	&27.64/0.7578\\
\cline{2-5} &
RED30 \cite{mao2016image} &32.94/0.9144	&29.61/0.8341	&27.86/0.7718\\
\cline{2-5} &
DnCNN\cite{zhang2017beyond} &33.03/0.9128	&29.81/0.8321	&28.04/0.7672\\
\cline{2-5} &
TNRD\cite{chen2016trainable} &32.51/0.9069	&29.43/0.8232	&27.66/0.7563\\
\cline{2-5} &
FDSR\cite{lu2018fast} &33.00/0.9042	&29.61/0.8179	&27.86/0.7500\\
\cline{2-5} &
SRCNN\cite{dong2015image} &32.42/0.9063 &29.28/0.8209 &27.49/0.7503\\
\cline{2-5} &
FSRCNN\cite{dong2016accelerating} &32.63/0.9088 &29.43/0.8242	&27.59/0.7535\\
\cline{2-5} &
RCN\cite{shi2017structure} &32.77/0.9109	&29.63/0.8269	&27.79/0.7594\\
\cline{2-5} &
VDSR\cite{kim2016accurate} &33.03/0.9124	&29.77/0.8314	&28.01/0.7674\\
\cline{2-5} &
DRCN\cite{kim2016deeply} &33.04/0.9118	&29.76/0.8311	&28.02/0.7670\\
\cline{2-5} &
CNF\cite{ren2017image} &33.38/0.9136 &29.90/0.8322 &28.15/0.7680\\
\cline{2-5} &
LapSRN\cite{lai2017deep} &33.08/0.9130	&29.63/0.8269 &28.19/0.7720\\
\cline{2-5} &
IDN\cite{hui2018fast} &33.30/0.9148 &29.99/0.8354 &28.25/0.7730\\
\cline{2-5} &
DRRN\cite{tai2017image} &33.23/0.9136	&29.96/0.8349	&28.21/0.7720\\
\cline{2-5} &
BTSRN\cite{fan2017balanced} &33.20/- &29.90/- &28.20/-\\
\cline{2-5} &
MemNet\cite{tai2017memnet}	&33.28/0.9142	&30.00/0.8350	&28.26/0.7723\\
\cline{2-5} &
CARN-M\cite{ahn2018fast} &33.26/0.9141	&30.08/0.8367	&28.42/0.7762\\
\cline{2-5} &
CARN\cite{ahn2018fast} &\textcolor{blue}{33.52/0.9166}	&\textcolor{blue}{30.29}/0.8407	&\textcolor{blue}{28.60/0.7806}\\
\cline{2-5} &
EEDS+\cite{wang2019end} &33.21/0.9151 &29.85/0.8339 &28.13/0.7698\\
\cline{2-5} &
DRFN\cite{yang2018drfn} &33.29/0.9142 &30.06/0.8366 &28.30/0.7737\\
\cline{2-5} &
MSDEPC\cite{liu2019single} &32.94/0.9111	&29.62/0.8279	&27.79/0.7581\\
\cline{2-5} &
CFSRCNN\cite{tian2020coarse}	&33.51/0.9165	&30.27/\textcolor{blue}{0.8410}	&28.57/0.7800\\
\cline{2-5} &
LESRCNN\cite{tian2020lightweight} &33.32/0.9148	&30.12/0.8380	&28.44/0.7772\\
\cline{2-5} &
LESRCNN-S\cite{tian2020lightweight}	&33.30/0.9145	&30.16/0.8384	&28.43/0.7776\\
\cline{2-5} &
ACNet\cite{tian2021asymmetric}	&33.41/0.9160	&30.19/0.8398	&28.46/0.7788\\
\cline{2-5} &
ACNet-B\cite{tian2021asymmetric} &33.32/0.9151 &30.15/0.8386	&28.41/0.7773\\
\cline{2-5} &
DIP-FKP\cite{liang2021flow} &27.06/0.7421 &26.27/0.6922 &25.65/0.6764\\
\cline{2-5} &
DIP-FKP + USRNet\cite{liang2021flow} &28.18/0.8088 &27.76/0.7750 &26.70/0.7383\\
\cline{2-5} &
KOALA\cite{kim2021koalanet} &30.35/0.8568 &- &27.20/0.7541\\
\cline{2-5} &
FALSR-B\cite{chu2021fast}   &33.29/0.9143 &- &-\\
\cline{2-5} &
FALSR-C\cite{chu2021fast}  &33.26/0.9140 &- &-\\
\cline{2-5} &
ESCN \cite{wang2017ensemble} &32.67/0.9093 &29.51/0.8264 &27.75/0.7611\\
\cline{2-5} &
HDN\cite{jiang2020hierarchical} &33.49/0.9150 &30.23/0.8400 &28.58/0.7810 \\
\cline{2-5} &
HGSRCNN (Ours)	&\textcolor{red}{33.56/0.9175}	&\textcolor{red}{30.32/0.8413}	&\textcolor{red}{28.62/0.7820}\\
\hline
\end{tabular}}
\label{tab:booktabs}
\end{table}
\subsection{Comparisons with state-of-the-arts}
To evaluate super-resolution effects of HGSRCNN from different angles, this paper conducts experiments in terms of quantitative and qualitative analysis. Specifically, quantitative analysis is used to test SR results containing PSNR, SSIM, run-time of restoring high-quality images, complexities, and perceptual quality of feature similarity index (FSIM) \cite{zhang2011fsim} of popular SR techniques, containing Bicubic \cite{sun2008image}, A+ \cite{timofte2014a+}, RFL \cite{schulter2015fast}, self-exemplars super-resolution (SelfEx) \cite{huang2015single}, a denoising CNN (DnCNN) \cite{zhang2017beyond}, the cascade of sparse coding based networks (CSCN) \cite{wang2015deep}, 30-layer residual encoder-decoder network (RED30) \cite{mao2016image}, trainable nonlinear reaction diffusion (TNRD) \cite{chen2016trainable}, fast dilated SR convolutional network (FDSR) \cite{lu2018fast}, a SR CNN (SRCNN) \cite{dong2015image}, fast super-resolution CNN (FSRCNN) \cite{dong2016accelerating}, residue context network (RCN) \cite{shi2017structure},  very deep SR network (VDSR) \cite{kim2016accurate}, deeply-recursive convolutional network (DRCN) \cite{kim2016deeply},  context-wise network fusion (CNF) \cite{ren2017image}, Laplacian super-resolution network (LapSRN) \cite{lai2017deep}, information distillation network (IDN) \cite{hui2018fast}, deep recursive residual network (DRRN) \cite{tai2017image}, balanced two-stage residual networks (BTSRN) \cite{fan2017balanced}, memory network (MemNet) \cite{tai2017memnet}, cascading residual network mobile (CARN-M)  \cite{ahn2018fast}, end-to-end deep and shallow network (EEDS+) \cite{wang2019end}, deep recurrent fusion network (DRFN) \cite{yang2018drfn}, multi-scale deep encoder-decoder with phase congruency (MSDEPC) \cite{liu2019single}, residual dense network (RDN)  \cite{zhang2018residual}, channel-wise and spatial feature modulation (CSFM) \cite{hu2019channel}, CFSRCNN  \cite{tian2020coarse},  lightweight enhanced SR CNN (LESRCNN) \cite{tian2020lightweight}, LESRCNN for varying scales (LESRCNN-S) \cite{tian2020lightweight}, asymmetric CNN (ACNet) \cite{tian2021asymmetric}, ACNet for blind SR (ACNet-B) \cite{tian2021asymmetric}, flow-based kernel prior (FKP) \cite{liang2021flow}, kernel-oriented adaptive local adjustment (KOALA) 
\cite{kim2021koalanet}, FALSR-B\cite{chu2021fast},  FALSR-C\cite{chu2021fast}, ensemble based sparse coding network (ESCN) \cite{wang2017ensemble},  hierarchical dense connection network (HDN) 
on four public datasets, i.e., Set5 \cite{bevilacqua2012low}, Set14 \cite{bevilacqua2012low}, B100 \cite{martin2001database} and U100 \cite{huang2015single} for different scales ($\times 2$, $\times 3$, $\times 4$), where HGSRCNN is obtained when control value is 0. Also, qualitative analysis is used to measure visual effects of different SR methods.

\begin{table}[t!]
\caption{Average PSNR/SSIM results of different SR techniques for different scales ($\times 2$, $\times 3$ and $\times 4$) on B100.}
\label{tab:1}
\centering
\scalebox{0.75}[0.75]{
\begin{tabular}{|c|c|c|c|c|}
\hline
\multirow{2}{*}{Dataset} &
\multirow{2}{*}{Methods} &
$\times 2$ & $\times 3$ & $\times 4$\\
\cline{3-5} & &PSNR/SSIM &PSNR/SSIM &PSNR/SSIM\\
\hline
\multirow{32}{*}{B100} &
Bicubic\cite{sun2008image}	&29.56/0.8431	&27.21/0.7385	&25.96/0.6675\\
\cline{2-5} &
A+\cite{timofte2014a+}	&31.21/0.8863	&28.29/0.7835	&26.82/0.7087\\
\cline{2-5} &
RFL\cite{schulter2015fast} &31.16/0.8840	&28.22/0.7806	&26.75/0.7054\\
\cline{2-5} &
SelfEx\cite{huang2015single}	&31.18/0.8855	&28.29/0.7840 &26.84/0.7106\\
\cline{2-5} &
CSCN\cite{wang2015deep} &31.40/0.8884	&28.50/0.7885 &27.03/0.7161\\
\cline{2-5} &
RED30\cite{mao2016image} &31.99/0.8974	&28.93/0.7994 &27.40/0.7290\\
\cline{2-5} &
DnCNN\cite{zhang2017beyond}	&31.90/0.8961	&28.85/0.7981	&27.29/0.7253\\
\cline{2-5} &
TNRD\cite{chen2016trainable} &31.40/0.8878	&28.50/0.7881	&27.00/0.7140\\
\cline{2-5} &
FDSR\cite{lu2018fast} &31.87/0.8847	&28.82/0.7797	&27.31/0.7031\\
\cline{2-5} &
SRCNN\cite{dong2015image}	&31.36/0.8879	&28.41/0.7863	&26.90/0.7101\\
\cline{2-5} &
FSRCNN\cite{dong2016accelerating} &31.53/0.8920	&28.53/0.7910	&26.98/0.7150\\
\cline{2-5} &
VDSR\cite{kim2016accurate} &31.90/0.8960	&28.82/0.7976	&27.29/0.7251\\
\cline{2-5} &
DRCN\cite{kim2016deeply} &31.85/0.8942	&28.80/0.7963	&27.23/0.7233\\
\cline{2-5} &
CNF\cite{ren2017image}	&31.91/0.8962	&28.82/0.7980	&27.32/0.7253\\
\cline{2-5} &
LapSRN\cite{lai2017deep}	&31.80/0.8950 &-	&27.32/0.7280\\
\cline{2-5} &
IDN\cite{hui2018fast} &32.08/\textcolor{blue}{0.8985} &28.95/0.8013 &27.41/0.7297\\
\cline{2-5} &
DRRN\cite{tai2017image} &32.05/0.8973 &28.95/0.8004 &27.38/0.7284\\
\cline{2-5} &
BTSRN\cite{fan2017balanced} &32.05/-    &28.97/- &27.47/- \\
\cline{2-5} &
MemNet\cite{tai2017memnet} &32.08/0.8978 &28.96/0.8001 &27.40/0.7281\\
\cline{2-5} &
CARN-M\cite{ahn2018fast}	&31.92/0.8960	&28.91/0.8000	&27.44/0.7304\\
\cline{2-5} &
CARN\cite{ahn2018fast}	&32.09/0.8978	&\textcolor{blue}{29.06}/0.8034	&\textcolor{blue}{27.58}/0.7349\\
\cline{2-5} &
EEDS+\cite{wang2019end} &31.95/0.8963 &28.88/\textcolor{red}{0.8054} &27.35/0.7263\\
\cline{2-5} &
DRFN\cite{yang2018drfn} &32.02/0.8979 &28.93/0.8010  &27.39/0.7293\\
\cline{2-5} &
MSDEPC\cite{liu2019single} &31.64/0.8961 &28.58/0.7918	&27.10/0.7193\\
\cline{2-5} &
CFSRCNN\cite{tian2020coarse} 	&\textcolor{blue}{32.11}/\textcolor{red}{0.8988}	&29.03/0.8035	&27.53/0.7333\\
\cline{2-5} &
LESRCNN\cite{tian2020lightweight} 	&31.95/0.8964	&28.91/0.8005	&27.45/0.7313\\
\cline{2-5} &
LESRCNN-S\cite{tian2020lightweight}	&31.95/0.8965	&28.94/0.8012	&27.47/0.7321\\
\cline{2-5} &
ACNet\cite{tian2021asymmetric}	&32.06/0.8978	&28.98/0.8023	&27.48/0.7326\\
\cline{2-5} &
ACNet-B\cite{tian2021asymmetric} &31.97/0.8970	&28.97/0.8016	&27.46/0.7316\\
\cline{2-5} &
DIP-FKP\cite{liang2021flow} &26.72/0.7089 & 25.96/0.6660 &25.15/0.6354\\
\cline{2-5} &
DIP-FKP + USRNet\cite{liang2021flow} &28.61/0.8206 &27.29/0.7484 &25.97/0.6902\\
\cline{2-5} &
KOALA\cite{kim2021koalanet} &29.70/0.8248 &- &26.97/0.7172\\
\cline{2-5} &
FALSR-B\cite{chu2021fast}   & 31.97/0.8967 &- &-\\
\cline{2-5} &
FALSR-C\cite{chu2021fast}  &31.96/0.8965 &- &-\\
\cline{2-5} &
ESCN \cite{wang2017ensemble} &31.54/0.8909 &28.58/0.7917 &27.11/0.7197\\
\cline{2-5} &
HDN\cite{jiang2020hierarchical} &32.03/0.8980 &28.96/0.8040 &27.53/\textcolor{red}{0.7370}\\
\cline{2-5} &
HGSRCNN (Ours)	&\textcolor{red}{32.12}/0.8984	&\textcolor{red}{29.09}/\textcolor{blue}{0.8042}	&\textcolor{red}{27.60}/\textcolor{blue}{0.7363}\\
\hline
\end{tabular}}
\label{tab:booktabs}
\end{table}

Quantitative analysis: Average PSNR and SSIM values of different SR methods can be obtained on four benchmark datasets, Set5, Set14, B100 and U100 as illustrated in TABLEs III-VI, where red and blue lines are symbolled as the best and second results for SISR, respectively. Specifically, we can see that HGSRCNN almost has obtained the best results for $ \times 2$, $ \times 3$ and $\times 4$ from TABLEs III-VI. In terms of small volume samples (i.e., Set5 and Set14), HGSRCNN is excellent in SISR. For instance, the HGSRCNN obtains an improvement of 0.06dB in PSNR and 0.0005 in SSIM on Set5 than that of the second CARN for $\times 3$  in TABLE III. Also, HGSRCNN achieves a prominent gain of 0.04dB and 0.0009 on Set14 than that of the second CARN in PSNR and  SSIM for $\times 2$ as shown in TABLE IV. In terms of big volume samples (i.e., B100 and U100), HGSRCNN also obtained remarkable results on SISR. For instance, HGSRCNN almost obtains the best effect for all the scales on U100 in SISR as illustrated in TABLE VI. Although the state-of-art RDN, CSFM and an image super-resolution feedback network (SRFBN) \cite{li2019feedback} are superior to HGSRCNN in PSNR and SSIM on a dataset with large volume samples (B100) for $\times 4$  as given in TABLE VII, they are faced to huger complexity and running time than that of HGSRCNN as reported in latter context. In a summary, these illustrations show that the proposed HGSRCNN obtains excellent performance to deal with LR images of different backgrounds.

\begin{table}[t!]
\caption{Average PSNR/SSIM results of different SR techniques for different scales ($\times 2$, $\times 3$ and $\times 4$) on U100.}
\label{tab:1}
\centering
\scalebox{0.85}[0.80]{
\begin{tabular}{|c|c|c|c|c|}
\hline
\multirow{2}{*}{Dataset} &
\multirow{2}{*}{Model} &
$\times 2$ & $\times 3$ & $\times 4$\\
\cline{3-5} & &PSNR/SSIM &PSNR/SSIM &PSNR/SSIM\\
\hline
\multirow{28}{*}{U100} &
Bicubic\cite{sun2008image}	&26.88/0.8403	&24.46/0.7349	&23.14/0.6577\\
\cline{2-5} &
A+\cite{timofte2014a+}	&29.20/0.8938	&26.03/0.7973	&24.32/0.7183\\
\cline{2-5} &
RFL\cite{schulter2015fast}	&29.11/0.8904	&25.86/0.7900	&24.19/0.7096\\
\cline{2-5} &
SelfEx\cite{huang2015single}	&29.54/0.8967	&26.44/0.8088 &24.79/0.7374\\
\cline{2-5} &
RED30\cite{mao2016image} &30.91/0.9159 &27.31/0.8303 &25.35/0.7587\\
\cline{2-5} &
DnCNN\cite{zhang2017beyond} &30.74/0.9139	&27.15/0.8276	&25.20/0.7521\\
\cline{2-5} &
TNRD\cite{chen2016trainable} &29.70/0.8994	&26.42/0.8076	&24.61/0.7291\\
\cline{2-5} &
FDSR\cite{lu2018fast} &30.91/0.9088	&27.23/0.8190	&25.27/0.7417\\
\cline{2-5} &
SRCNN\cite{dong2015image} &29.50/0.8946	&26.24/0.7989	&24.52/0.7221\\
\cline{2-5} &
FSRCNN\cite{dong2016accelerating} &29.88/0.9020	&26.43/0.8080	&24.62/0.7280\\
\cline{2-5} &
VDSR\cite{kim2016accurate} &30.76/0.9140	&27.14/0.8279	&25.18/0.7524\\
\cline{2-5} &
DRCN\cite{kim2016deeply} &30.75/0.9133	&27.15/0.8276	&25.14/0.7510\\
\cline{2-5} &
LapSRN\cite{lai2017deep} &30.41/0.9100	&-	&25.21/0.7560\\
\cline{2-5} &
IDN\cite{hui2018fast} &31.27/0.9196 &27.42/0.8359 &25.41/0.7632\\
\cline{2-5} &
DRRN\cite{tai2017image} &31.23/0.9188	&27.53/0.8378	&25.44/0.7638\\
\cline{2-5} &
BTSRN\cite{fan2017balanced} &31.63/- &27.75/- &25.74- \\
\cline{2-5} &
MemNet\cite{tai2017memnet}	&31.31/0.9195	&27.56/0.8376	&25.50/0.7630\\
\cline{2-5} &
CARN-M\cite{ahn2018fast}	&30.83/0.9233	&26.86/0.8263	
&25.63/0.7688\\
\cline{2-5} &
CARN\cite{ahn2018fast} &31.51/\textcolor{red}{0.9312} &27.38/0.8404	&26.07/0.7837\\
\cline{2-5} &
DRFN\cite{yang2018drfn} &31.08/0.9179 &27.43/0.8359 &25.45/0.7629\\
\cline{2-5} &
CFSRCNN\cite{tian2020coarse} &\textcolor{blue}{32.07}/0.9273	&\textcolor{blue}{28.04/0.8496}	&26.03/0.7824\\
\cline{2-5} &
LESRCNN\cite{tian2020lightweight}	&31.45/0.9206	&27.70/0.8415	&25.77/0.7732\\
\cline{2-5} &
LESRCNN-S\cite{tian2020lightweight}	&31.45/0.9207	&27.76/0.8424	&25.78/0.7739\\
\cline{2-5} &
ACNet\cite{tian2021asymmetric} 	&31.79/0.9245	&27.97/0.8482	&25.93/0.7798\\
\cline{2-5} &
ACNet-B\cite{tian2021asymmetric} &31.57/0.9222	&27.88/0.8447	&25.86/0.7760\\
\cline{2-5} &
DIP-FKP\cite{liang2021flow} & 24.33/0.7069 &23.47/0.6588 & 22.89/0.6327\\
\cline{2-5} &
DIP-FKP + USRNet\cite{liang2021flow} &26.46/0.8203 &24.84/0.7510 &23.89/0.7078\\
\cline{2-5} &
KOALA\cite{kim2021koalanet} &27.19/0.8318 &- &24.71/0.7427 \\
\cline{2-5} &
FALSR-B\cite{chu2021fast}   &31.28/0.9191 &- &-\\
\cline{2-5} &
FALSR-C\cite{chu2021fast}  &31.24/0.9187 &- &-\\
\cline{2-5} &
HDN\cite{jiang2020hierarchical} &31.87/0.9250  &27.93/0.8490 &\textcolor{blue}{26.09/0.7870}\\
\cline{2-5} &
HGSRCNN (Ours) &\textcolor{red}{32.21}/\textcolor{blue}{0.9292} &\textcolor{red}{28.29/0.8546} &\textcolor{red}{26.27/0.7908}\\
\hline
\end{tabular}}
\label{tab:booktabs}
\end{table}

It is known that digital devices have demands for execution time and complexity  \cite{tian2020coarse,martin2001database}. According to that, we use eight popular SR methods, i.e., VDSR, DRRN, MemNet, RDN, SRFBN, CARN-M, CFSRCNN and ACNet as comparative methods to restore high-quality images with $256 \times 256$, $512 \times 512$ and $1024 \times 1024$ on $\times 2$ to test running time of these methods. As described in TABLE VIII, we can see that HGSRCNN achieves execution fast in SISR. That is, HGSRCNN takes the run-time to $4.58\%$ of popular RDN, $4.27\%$ of SRFBN in predicting a HR image with size of $1024 \times 1024$. In terms of complexity, we exploit VDSR and DnCNN, DRCN, MemNet, CARN-M, CARN, CSFM, RDN, SRFBN, ACNet and HGSRCNN to conduct experiments for measuring their complexities. Specifically, the number of parameters and flops \cite{tian2020attention} of training a SR model are used to express as complexity of  computational cost and memory consumption for predicting SR images of size $162 \times 162$. As given in TABLE IX, HGSRCNN only takes the number of parameters to $9.9\%$ of 134-layer RDN and $16.9\%$ of 384-layer CSFM to obtain approximative SR results. Besides, TABLE IX reports that HGSRCNN only takes $10.40\%$ of RDN and $17.70\%$ of CSFM in flops. Thus, HGSRCNN is a useful SR tool in terms of PSNR, SSIM, run-time and complexity. 
 
To comprehensively evaluate SISR performance of the proposed HGSRCNN, we use FSIM values of different methods to test their visual effects in terms of perception. TABLE X proves that the proposed HGSRCNN obtained the highest values than these of CFSRCNN and ACNet on B100 for three different scales (i.e., $\times 2$, $\times 3$  and $ \times 4$). According to mentioned illustrations, we can see that the proposed HGSRCNN is very effective in quantitative analysis for SISR. 

\begin{table}[t!]
\caption{Average PSNR/SSIM values of different SR methods for $\times 4$ on B100.}
\label{tab:1}
\centering
\scalebox{0.75}[0.75]{
\begin{tabular}{|c|c|c|}
\hline
Methods	&PSNR(dB)	&SSIM\\
\hline
RDN	\cite{zhang2018residual} &27.72	&0.7419\\
\hline
CSFM \cite{hu2019channel}	&27.76	&0.7432\\
\hline
SRFBN \cite{li2019feedback}	&27.72	&0.7409\\
\hline
CFSRCNN	\cite{tian2020coarse} &27.53	&0.7333\\
\hline
HGSRCNN (Ours) &27.60 &0.7363\\
\hline
\end{tabular}}
\label{tab:booktabs}
\end{table}
\begin{table}[t!]
\caption{Running time (seconds) of different SR methods on predicting HR images of sizes $256\times256$, $512\times512$ and $1024\times1024$ for $\times2$.}
\label{tab:1}
\centering
\scalebox{0.85}[0.86]{
\begin{tabular}{|c|c|c|c|}
\hline
\multicolumn{4}{|c|}{Single Image Super-Resolution} \\
\hline
Size &$256 \times 256$ &$512 \times 512$ &$1024 \times 1024$\\
\hline
VDSR\cite{kim2016accurate} &0.0172	&0.0575  &0.2126\\
\hline
DRRN\cite{tai2017image}  &3.063 	&8.050  &25.23\\
\hline
MemNet\cite{tai2017memnet} &0.8774		&3.605   &14.69\\
\hline
RDN \cite{zhang2018residual} &0.0553 &0.2232 &0.9124\\
\hline
SRFBN \cite{li2019feedback} &0.0761 &0.2508 &0.9787\\
\hline
CARN-M\cite{ahn2018fast} &0.0159 &0.0199 &0.0320\\
\hline
CFSRCNN\cite{tian2020coarse} &0.0153	&0.0184	&0.0298\\
\hline
ACNet\cite{tian2021asymmetric} &0.0166	&0.0195	&0.0315\\
\hline
HGSRCNN (Ours)	&0.0234	&0.0337	&0.0418\\
\hline
\end{tabular}}
\label{tab:booktabs}
\end{table}
\begin{table}[t!]
\caption{Complexities of different SR methods for $\times 2$.}
\label{tab:1}
\centering
\scalebox{0.6}[0.6]{
\begin{tabular}{|c|c|c|}
\hline
Methods	&Parameters	&Flops\\
\hline
VDSR\cite{kim2016accurate}	&665K	&17.45G\\
\hline
DnCNN\cite{zhang2017beyond}	&556K	&14.59G\\
\hline
DRCN\cite{kim2016deeply}	&1,774K	&46.56G\\
\hline
MemNet\cite{tai2017memnet}	&677K	&17.77G\\
\hline
CARN-M\cite{ahn2018fast} &412K	&3.46G\\
\hline
CARN\cite{ahn2018fast} 	&1,592K	&11.21G\\
\hline
CSFM \cite{hu2019channel} &12,841K &85.01G\\
\hline
RDN \cite{zhang2018residual} &21,937K &144.69G\\
\hline
SRFBN \cite{li2019feedback} &3,631K &23.86G\\
\hline
ACNet \cite{tian2021asymmetric} &1,283K &14.25G\\
\hline
HGSRCNN (Ours) &2,178K &15.05G\\
\hline
\end{tabular}}
\label{tab:booktabs}
\end{table}
\begin{table}[htbp!]
\caption{FSIM values of SR methods for $\times 2$, $\times 3$ and $\times 4$ on B100.}
\label{tab:1}
\centering
\scalebox{0.89}[0.90]{
\begin{tabular}{|c|c|c|c|c|}
\hline
\multirow{1}{*}{Dataset} &
\multirow{1}{*}{Methods} &
$\times 2$ & $\times 3$ & $\times 4$\\
\hline
\multirow{6}{*}{B100} &
A+\cite{timofte2014a+}	&0.9851	&0.9734	&0.9592\\
\cline{2-5} &
SelfEx\cite{huang2015single} &0.9976	&0.9894	&0.9760\\
\cline{2-5} &
SRCNN\cite{dong2015image} &0.9974	&0.9882	&0.9712\\
\cline{2-5} &
CARN-M\cite{ahn2018fast} &0.9979	&0.9898	&0.9765\\
\cline{2-5} &
LESRCNN\cite{tian2020lightweight} &0.9979	&0.9903	&0.9774\\
\cline{2-5} &
CFSRCNN\cite{tian2020coarse}	&0.9980	&0.9905	&0.9776\\
\cline{2-5} &
ACNet\cite{tian2021asymmetric}	&0.9980	&0.9905	&0.9777\\
\cline{2-5} &
HGSRCNN (Ours)	&0.9980	&0.9906	&0.9785\\
\hline
\end{tabular}}
\label{tab:booktabs}
\end{table}
\begin{figure}[!htbp]
\centering
\subfloat{\includegraphics[width=3.5in]{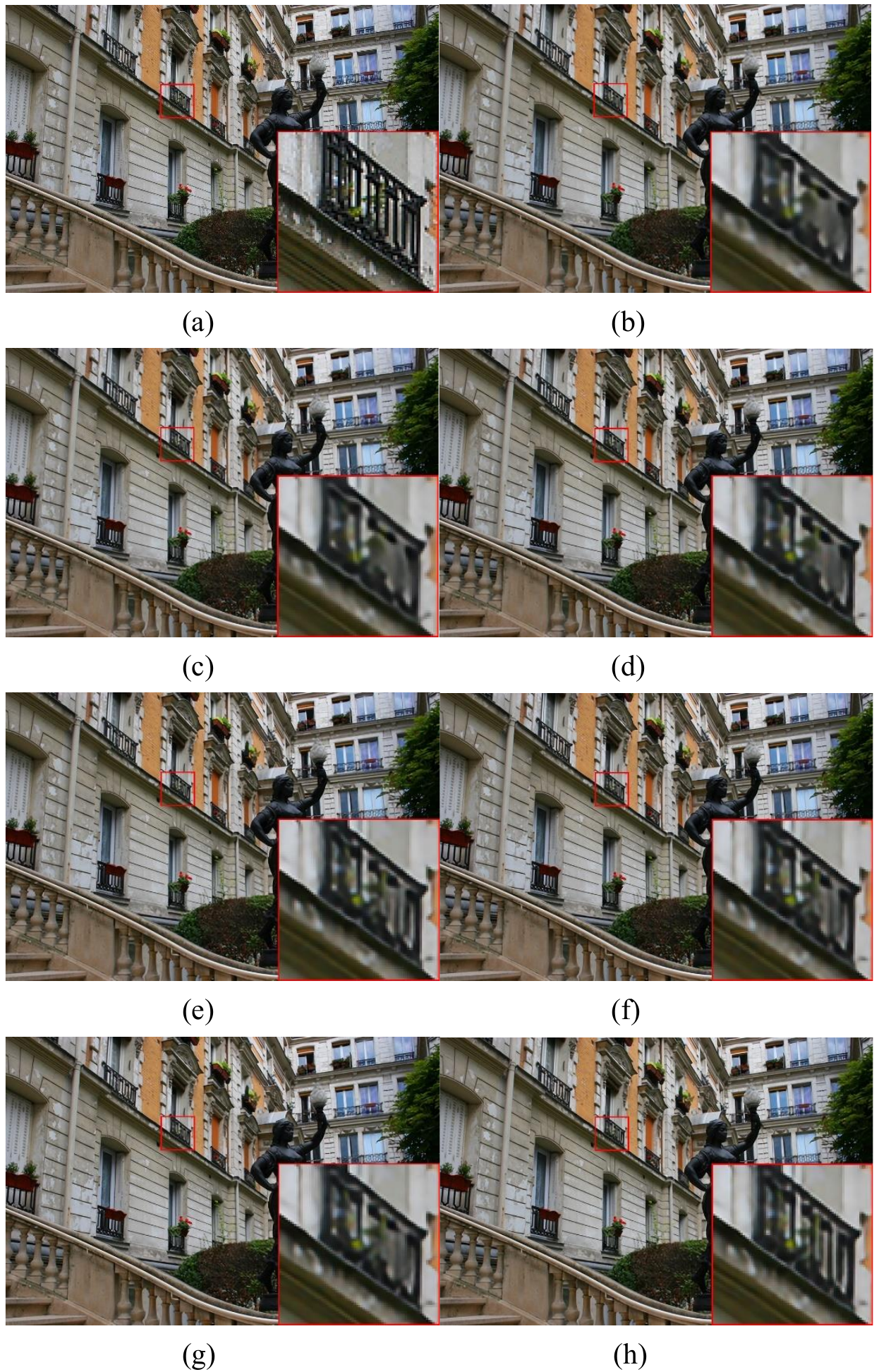}
\label{fig_second_case}}
\caption{Visual effect of different methods for $\times 3$ on U100 as follows. (a) HR image, (b) VDSR, (c) DRCN, (d) CARN-M, (e) LESRCNN,
(f) CFSRCNN, (g) ACNet and (h) HGSRCNN (Ours).}
\label{fig:7}
\end{figure}

\begin{figure*}[!htbp]
\centering
\subfloat{\includegraphics[width=4.5in]{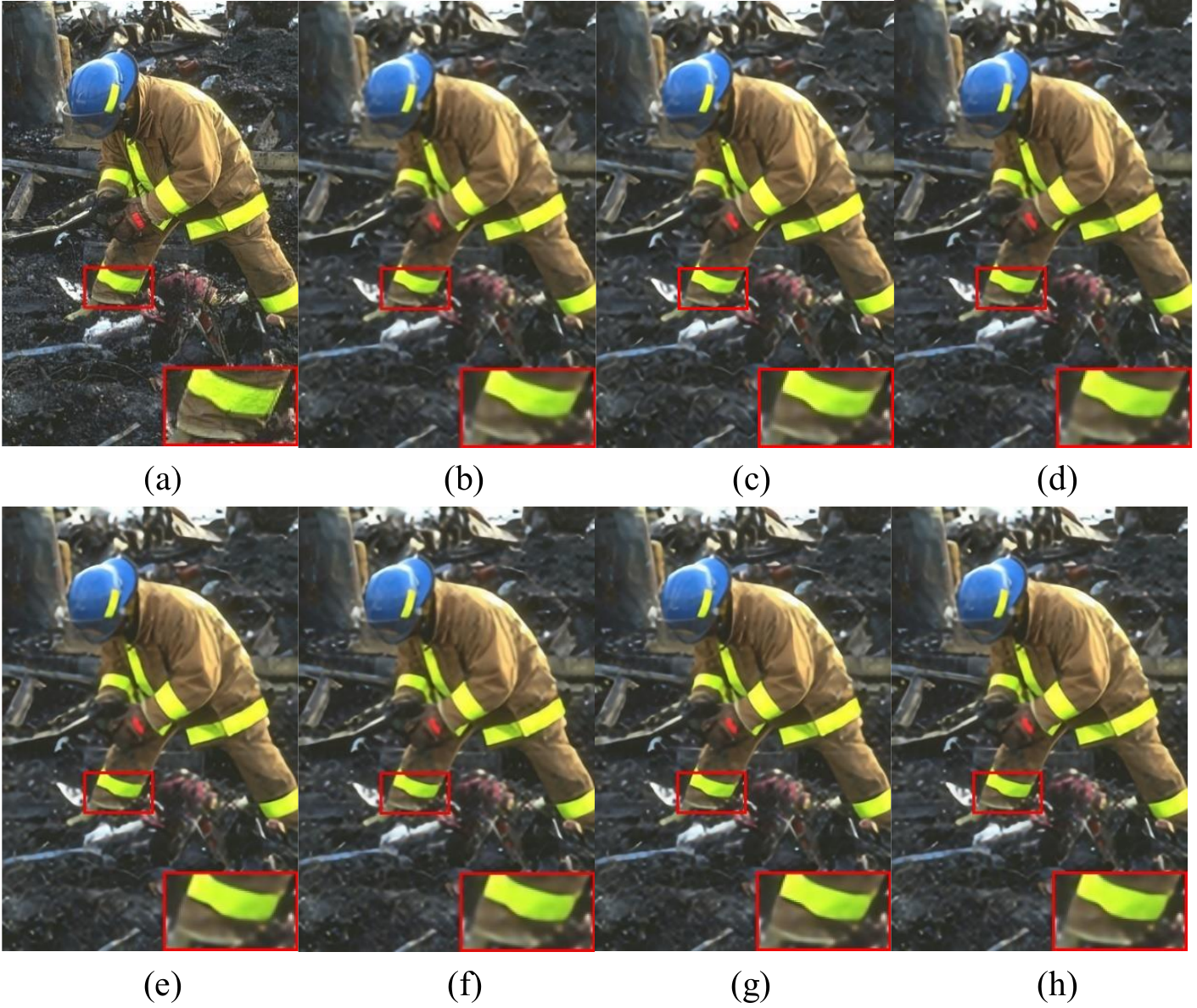}
\label{fig_second_case}}
\caption{Visual effect of different methods for $\times 4$ on B100 as follows. 
(a) HR image, (b) VDSR, (c) DRCN, (d) CARN-M, (e) LESRCNN,
 (f) CFSRCNN, (g) ACNet and (h) HGSRCNN (Ours).}
\label{fig:7}
\end{figure*}
Qualitative analysis: To test visual results of the proposed HGSRCNN, we choose six popular methods (i.e., VDSR, DRCN, CRAN-M, LESRCNN, CFSRCNN and ACNet) on U100 and B100 to conduct predicted high-quality images. To easier observe detailed information of constructed SR images from different methods, one area of the predicted image is amplified as an observation area. The observation area is clearer, which implies its corresponding SR method has better performance. Figs. 3-4. point out that marked regions of the HGSRCNN are clearer than these of other SR methods. In other words, the proposed HGSRCNN outperforms other methods for SISR. According to quantitative analysis and qualitative analysis, we can see that the proposed HGSRCNN is beneficial to SISR on digital devices. 

This paper has the following contributions.

(1)	The proposed 52-layer HGSRCNN uses heterogeneous architectures and refinement blocks to enhance internal and external interactions of different channels both in parallel and serial ways for obtaining richer low-frequency structure information of different types, which is very suitable to SISR in complex scenes.

(2)	A multi-level enhancement mechanism guides a CNN to implement a symmetric architecture for progressively facilitating structural information in SISR.

(3) The designed HGSRCNN obtains competitive execution speed for SISR. That is, it only takes the run-time to $4.58\%$ of RDN and $4.27\%$ of SRFBN in restoring a high-quality image with $1024 \times 1024$.
\section{Conclusion}
In this paper, we propose a heterogeneous group SR CNN (HGSRCNN). The HGSRCNN uses heterogeneous architectures in a parallel way to enhance internal and external relations of different channels for facilitating riches low-frequency structure information. Taking effects of obtained redundant features into consideration, a refinement block with signal enhancements in a serial way is conducted to filter useless information. To prevent loss of original information, a multi-level enhancement mechanism guides a CNN to implement a symmetric architecture for promoting expressive ability of HGSRCNN. Besides, a parallel up-sampling mechanism is developed to train a blind SR model. A lot of experiments are conducted on four benchmark datasets to prove the effectiveness of the proposed HGSRCNN in terms of SISR results, SISR efficiency, complexity and visual effects.





\ifCLASSOPTIONcaptionsoff
  \newpage
\fi



%

\bibliographystyle{IEEEtran}
\bibliography{IMSC_AGL}

…

\end{document}